\documentstyle[12pt,aaspp4,psfig]{article}
\def\sci#1#2{#1$\times$10$^{#2}$}

\received{1996 May 7}
\lefthead{Ford \& Band}
\righthead{Extrapolations of BATSE Spectra}

\begin{document}

\title {Extrapolations of BATSE Gamma-Ray Burst Spectra \\ 
to the Optical-UV Band}
\author{Lyle A. Ford\altaffilmark{1} and David L. Band\altaffilmark{2}}
\affil{Center for Astrophysics and Space Sciences 0111,
University of California at San Diego,\\ 
La Jolla, CA 92093-0111}
\altaffiltext{1}{Present address:  Department of Physics and Astronomy, 
University of Wisconsin, Eau Claire, P. O. Box 4004, Eau Claire, WI  54702;
ford@miranda.uwec.edu}
\altaffiltext{2}{dband@ucsd.edu}
\centerline{To appear in {\it The Astrophysical Journal}}
\centerline{\it Received 1996 May 7; accepted 1996 July 10}
\begin{abstract}
Many gamma-ray burst counterpart searches are being conducted in the optical-UV
band. To both predict detectability and understand the meaning of any
detections or upper limits, we extrapolate gamma-ray spectra from 54 bright
gamma-ray bursts to optical-UV energies. We assume optical emission is
concurrent with gamma-ray emission and do not consider quiescent or fading
counterparts. We find that the spectrum must be steeper (greater flux at low
energy) than a simple extrapolation of the gamma-ray spectrum for more than one
simultaneous optical flash to be observable per year by current searches. 
\end{abstract}
\section{Introduction}
Despite years of study and volumes of data, gamma-ray bursts (GRBs) remain
among the most mysterious cosmic phenomena. Perhaps the greatest impediment to
theoretical progress is that the source of GRBs is unknown. Although it is
widely thought that neutron stars are involved, there is no consensus on even
the broadest details of the physical processes involved. The main reason for
this is that the distance to GRBs is not known.  Many have interpreted the
observations of GRB isotropy and spatial inhomogeneity (Fishman et al.\ 1994)
by the {\it Compton Gamma Ray Observatory}'s Burst and Transient Source
Experiment (BATSE) as indicating that bursts are at cosmological distances.
Attempts have been made to verify this conclusion with BATSE data (Davis et
al.\ 1994; Norris et al.\ 1994) but the claims are controversial (Band 1994;
Brainerd 1994) and the possibility of a Galactic origin has not been ruled out
(c.f., Podsiadlowski, Rees, \& Ruderman 1995). Uncertainty in the distance to
GRBs leads to uncertainty in the energy required to generate observed GRBs and
hence, confusion about the source physics. If the distance scale to GRBs can be
determined, it will be the greatest step yet taken toward understanding these
enigmatic events. 

An excellent way to establish the GRB distance scale is to discover
counterparts at other wavelengths. Counterparts could also reveal valuable
information about burst sources and environments which would have an enormous
impact on our physical understanding of these events. For example, if GRBs were
found only in a particular environment, then properties unique to that
environment are likely to be responsible for the burst. Many groups (see
reviews by Schaefer 1994 and Greiner 1995) have attempted to find counterparts
which flare along with the burst, are observable in quiescence, or appear as
afterglows. To date, convincing counterparts have been found only for soft
gamma repeaters, which many believe to be unrelated to GRBs (although this
opinion is not universal; see Rothschild \& Lingenfelter 1996). Several
theories have been been proposed which predict emission at wavelengths other
than gamma-rays (a number of them are listed in Schaefer 1994) although the
guidance they provide is slim considering that many were advanced at a time
when bursts were thought to originate at distances $\lesssim 1$~kpc, much
closer than is now believed. 

Since we have no clear expectations as to the appearance of a burst source in
quiescence or immediately after the gamma-ray event, we assume in this work
that all optical emission occurs at the same time as the high energy emission
and is the low energy tail of the gamma ray spectrum.  Just as we extrapolate
observed X-ray and gamma-ray emission to the optical-UV band, some model high
energy spectra can be extrapolated to lower energy.  Katz (1994) and Tavani
(1996a,b) proposed a model where particles in a relativistic shocked plasma
with an equilibrium energy distribution radiate a synchrotron spectrum which
can extend to optical energies. M\'esz\'aros \& Rees (1993) modeled the
multiwavelength emissions of a fireball which interacts with the surrounding
medium, resulting in a reverse shock, with emission from a number of regions. 
A variety of different high energy and optical-UV spectra are possible,
depending on the physical parameters (e.g., densities and magnetic field); for
some parameters an optical synchrotron spectrum similar to the Katz and Tavani
models is emitted, while for others the optical emission can be greater, or
self-absorbed. 

Several models predict concurrent low energy emission based on the reprocessing
of gamma-rays in a neutron star magnetosphere (Hartmann, Woosley, \& Arons
1988), an accretion disk (Epstein 1985) or a stellar companion (London \&
Cominsky 1983).  These models all assume nearby ($\sim 100$ pc) sources and
consequently much smaller source energies ($\lesssim 10^{38}$~ergs) than is
required for bursts in the Galactic halo or at cosmological distances ($\gg
10^{41}$~ergs); it is unclear what effect such a large energy release will have
on the reprocessing region. While these reprocessing models predict optical
re-emission while the burst is still in progress, the re-emitted flux should be
proportional to the total gamma ray energy flux and not to an extrapolation of
the gamma ray spectrum. 

Lacking compelling theoretical expectations, we take an empirical approach to
determine how bright simultaneous counterpart emission may be by extrapolating
spectra observed by BATSE to optical and ultraviolet energies (\S 2). Several
experiments capable of detecting simultaneous optical emission from a GRB are
described (\S 3) and the implications of our results for these experiments are
discussed (\S 4). 
\section{Analysis} 
We base our extrapolated fluxes on spectral fits to 54 bright GRBs from the
first 13 months of BATSE operation (Band et al.\ 1993). The bursts were
selected with a brightness criterion using the peak count rate in the
$50-300$~keV energy band. Most bursts which met the intensity requirement were
included in the sample, but there were a few exceptions. Bright bursts shorter
than 1s were left out of the sample while some dim bursts with interesting time
structure were included; the number in both cases was small. Although the
sample is not statistically complete, it is representative of bright bursts. 

In this work, we ignore interstellar absorption. While this effect will render
sources which must be observed through the Galactic plane invisible in the
optical-UV band, it is far less important above the plane. Near the Galactic
poles, absorption typically dims sources by $\Delta m_V \sim 0.3$ (Mihalas \&
Binney 1981, p.~181). The exact amount of absorption depends on the amount of
interstellar matter along the line of sight, something which is not a simple
function of elevation angle above the plane because of the clumpy nature of the
interstellar medium. Given the large range of magnitudes which results from our
extrapolation of the spectrum over four energy decades under very simple
assumptions, the additional complexity of modeling interstellar absorption of
$\Delta m_V \sim 0.3-1$ off the Galactic plane is not warranted. 

Band et al.\ (1993) fitted spectra averaged over the entire burst to a four
parameter function of the form 
\begin{equation}
N_E(E)~\biggl({{\rm photons}\over\hbox{keV-s-cm}^2}\biggr)=
\cases{A\biggl({E\over{100 {\rm ~keV}}}\biggr)^\alpha
e^{-E/E_0},&$E \le (\alpha-\beta)E_0$\cr
A^\prime\biggl({E\over{100 {\rm ~keV}}}\biggr)^\beta,&$E>(\alpha-\beta)E_0$\cr}
\end{equation}
where $A,~\alpha,~\beta,$ and $E_0$ were fitted to the observed spectra and
$A^\prime$ was chosen to make the function continuously differentiable
everywhere. The energy range covered by the fit varied, with typical lower and
upper energy cutoffs of $10-30$~keV and $1-3$~MeV, respectively. This is a
differential photon number spectrum which must be multiplied by energy,
$F_E(E)=N_E(E)E$, to get the energy flux. Note that the spectral indices in
eq.~(1) do not have an explicit minus sign. 

Here we extrapolate the fitted spectra down to 1~keV where they break to a
power law with a different spectral index which describes the optical through
soft X-ray spectra. The index of the low energy extrapolation is varied from
$-4$ to $4$ in integer steps. In addition, we include an energy index of 1/3,
which is physically interesting, as we discuss below.  Finally, a low energy
power law spectrum of index $\alpha+1$, an unmodified continuation of the gamma
ray spectrum, is extrapolated to the optical range. 

Discussions of counterpart emission are conventionally based on burst fluence;
expected magnitudes are presented as if all the emission occurred in one second
(Schaefer 1981). Although this is not consistent with the assumption of
strictly simultaneous optical emission, it is a useful exposure-time
independent standardization. We calculate fluence based magnitudes in the
energy bands of the U, B, and V filters and the $5-7$~eV bandpass of the UV
camera array on the {\it High Energy Transient Explorer} ({\it HETE}, discussed
in \S 3). For the first three bands, we convolved spectra with the filter
response functions (Budding 1993, p.~46) and converted the fluxes to
magnitudes. The {\it HETE} bandpass is not a standard wavelength band so we
integrate the flux over the $5-7$~eV band.  Expected magnitudes for a one
second exposure are given in Table~1 and distributions for the $\alpha+1$ and
1/3 extrapolations are shown in Figures 1 and~2 (the distributions for the
integral indices are identical to the distribution for 1/3, but with the
magnitude axis shifted).  The true differential distribution is a convolution
of the intensity distribution and the distribution of spectral shapes.  Since
the burst intensity distribution increases with decreasing intensity (perhaps
until the spatial inhomogeneity becomes particularly severe), we expect the
optical magnitude distribution to increase with magnitude.  However, our
calculated optical distributions are based on a sample of bright bursts, and
therefore the distributions are truncated on the large magnitude (low optical
flux) side.  If the optical searches can reach magnitudes larger than or
comparable to the peaks of the distributions in Figures 1 and~2, then the
number of bursts which these searches may detect will be underestimated. 
However, as will be demonstrated below, this is not an issue for physically
interesting extrapolations. 

\begin{figure}
\psfig{file=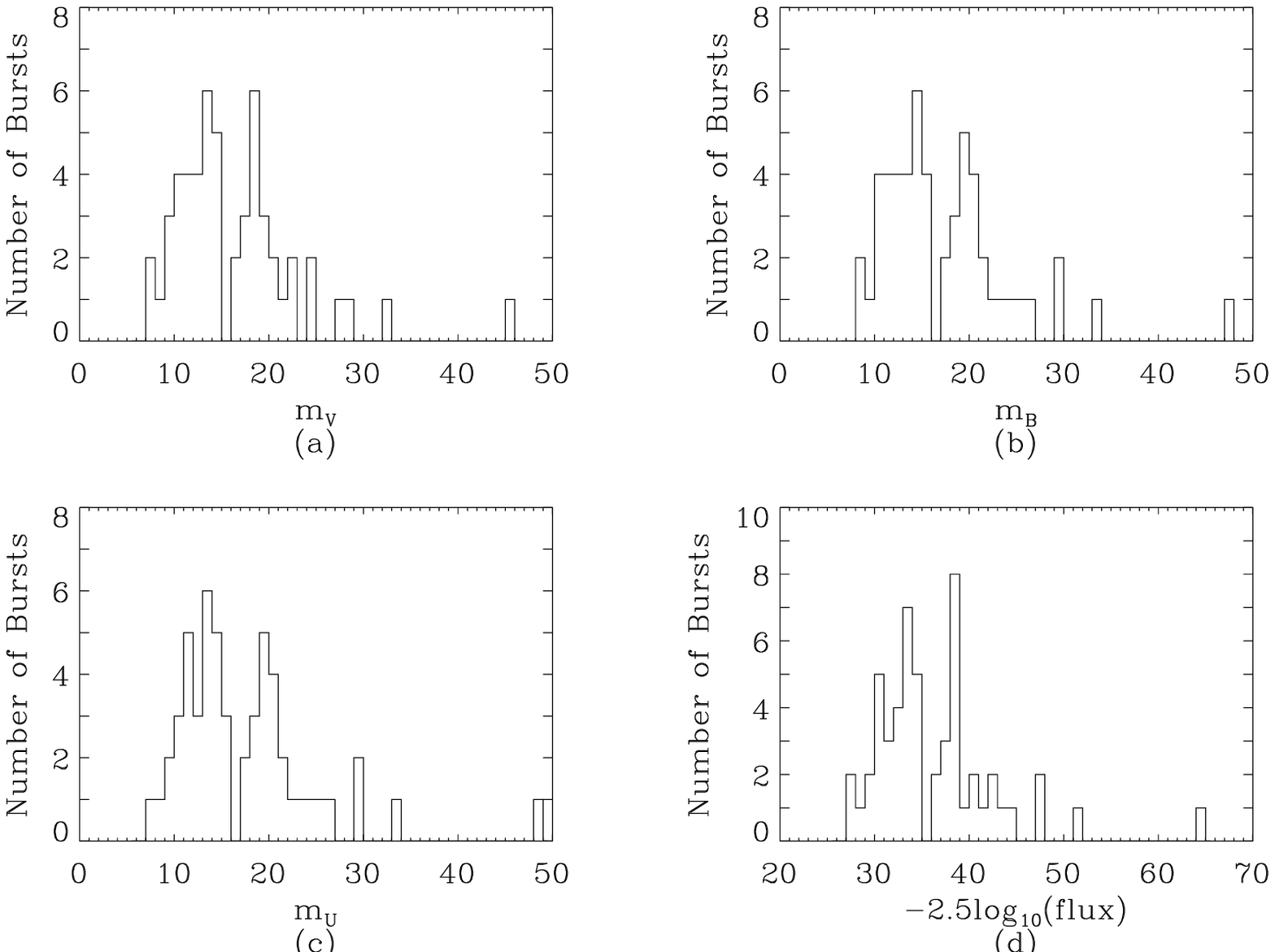}
\caption{The number of bursts at a given fluence-based magnitude (entire
burst occurs in one second) assuming the flux extrapolates as a power law of
index $\alpha +1$ from the gamma-rays and a one second exposure. Panels (a)-(d)
are the V, B, U, and $5-7$~eV ({\it HETE}) bands, respectively. Note: Only
bright bursts are included. If all BATSE bursts had been included in the
sample, this differential distribution would continue to rise as the magnitude
increased, not peak as shown here.} 
\end{figure}
\begin{figure}
\psfig{file=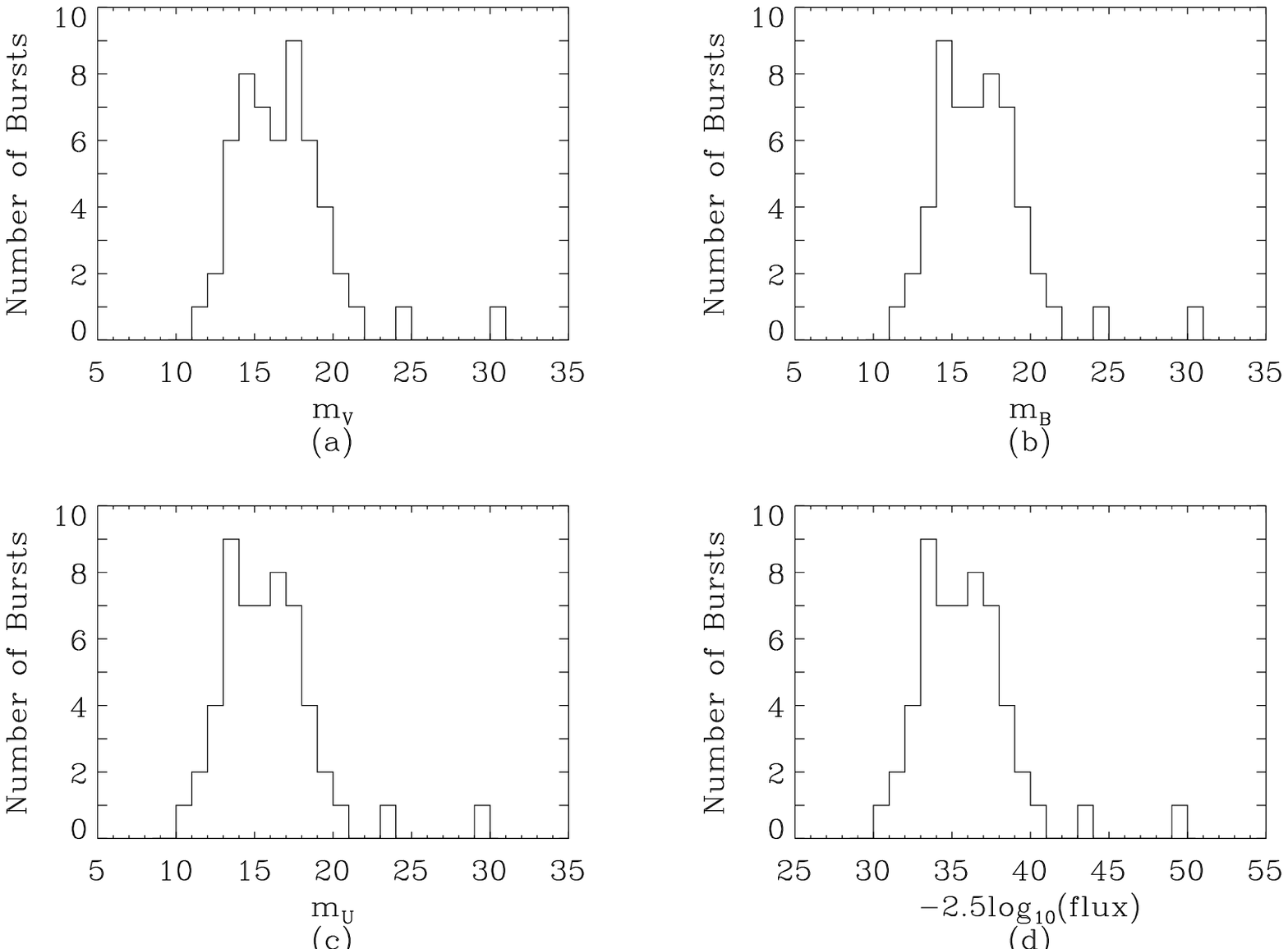}
\caption{Same as Figure 1, but for an energy index of 1/3.}
\end{figure}

Although fluence-based measures of brightness are the standard, flux-based
measures are more meaningful when detector exposure durations are comparable to
the burst duration (note that we assume the optical emission has the same
duration as the gamma-ray emission). This distinction is unimportant when the
exposure time is longer than the burst since the entire burst fluence will be
observed.  Indeed, the search for optical transients began with the inspection
of photographic plates which had exposure times much longer than the typical
burst duration. If, however, the exposure time is shorter than the burst, then
only a fraction of the total flux is observed and fluence-based measures will
overestimate the brightness of a burst. 

Another concern when dealing with real experiments is spectral response. All
the instruments we consider except {\it HETE} are ground based and sensitive
between $\sim 4000-7000$\AA. Although this energy band is very wide, the
convention is to define sensitivities in terms of V magnitudes. Therefore, we
convolve spectra with the response function of the V-filter and convert the
flux to magnitude for comparison. This will introduce slight errors in the
comparison of observability because the GRB spectra are assumed to be power
laws while the sensitivities of the instruments were determined using stars,
which have black body continua. These errors are not large however and a proper
treatment would require detailed knowledge of the complicated response
functions of the various experiments. 

For each instrument we calculated the expected observed flux within each
experiment's exposure time. In all cases, the detector response was assumed to
be linear (no saturation). If a burst was shorter than the exposure time, the
entire flux of the burst was averaged over the exposure time (equivalent to the
fluence-based magnitude with dilution for exposure times $>1$s). If the burst
was longer, we used the average flux over the exposure time. This process can
be expressed mathematically as 
\begin{equation}
m_V=2.5\log (Tf_V/f_0),
\end{equation}
where $f_V$ is the V band extrapolated flux and $f_0$ is the V band flux of a
magnitude zero object (3.08$\times 10^{-9}$ ergs/cm$^2$-s---Budding 1993,
p.~46). $T$ is defined as 
\begin{equation}
T=\cases{1 & $t_{exp} < t_{dur}$ \cr
t_{dur}/t_{exp} & $t_{exp} > t_{dur}$, }
\end{equation}
where $t_{exp}$ is the instrument exposure time and $t_{dur}$ is the burst
duration. The case $t_{exp} < t_{dur}$ is only an approximation to the true
instantaneous flux because intensity and spectra vary within a burst (Ford et
al. 1995). Although the spectral changes are not as radical as intensity
fluctuations, spectral variations are amplified by the four decade
extrapolation from the lowest energy gamma-rays included in the original
spectral fits. 

Better sensitivity can be obtained for the short exposure instruments by adding
subsequent images together. We did not consider this effect since approximately
one magnitude is gained for a typical burst. This boost in sensitivity is
likely to be more than compensated by our neglect of slew times for telescopes
which respond to a burst trigger since typically the brightest emission occurs
early in the burst. 
\section{Instrumentation}
Instruments which can detect simultaneous optical emission from GRBs can be
grouped into two categories. The first has large fields-of-view and long
exposure times ($>30$~min). These telescopes were designed for meteor patrols,
wide field imaging or variable star patrols, but their large fields-of-view
make them well suited for GRB detection. Many of these instruments are outlined
in Greiner et al.\ (1994). Those most likely to image a field containing a GRB
are sky patrols from Sonneberg and Ond\v rejov (the Odessa and Dushanbe sky
patrols have been suspended because of economic difficulties in Ukraine and
Tadshikistan---J.~Greiner 1996, private communication).  Although the patrol
plates have a variety of sensitivities and exposure times, we concentrate on
sky patrol plates from Ond\v rejov because its field-of-view is particularly
wide and it has three coincidences with bursts in the Band et al.\ (1993)
sample (Greiner et al.\ 1994). 

The second class of instruments we consider is characterized by short exposure
times. These instruments are much more sensitive to GRBs, which seldom last
more than a few minutes (the longest burst in our sample is 130s). We consider
four experiments here: the Explosive Transient Camera (ETC, Vanderspek et al.\
1994), the Gamma-Ray Optical Counterpart Search Experiment (GROCSE, Akerlof et
al.\ 1994), the Gamma-ray To Optical Transient Experiment (GTOTE, S. Barthelmy
1995, private communication), and the Ultraviolet Transient Camera Array on the
{\it HETE} spacecraft (Ricker et al.\ 1992; Vanderspek et al.\ 1995). All of
these instruments monitor a portion of the sky but the ETC, GROCSE, and GTOTE
can slew to preliminary BATSE positions distributed on the BACODINE network
(Barthelmy et al.\ 1994). BACODINE requires up to five seconds to localize a
GRB and another few seconds to distribute the coordinates on the network. After
that, the telescope must slew to the proper position. These delays will often
result in bright portions of the burst being missed by the telescope. We ignore
these concerns in this paper so our extrapolated brightness may be somewhat
optimistic. Details of all the instruments mentioned here are presented in
Table~2. 
\section{Discussion}
The flux-based magnitudes of bursts expected under our assumptions are given in
Table~3. Figures 3a and~b show the V-band brightness distribution of our sample
for extrapolations with energy indices of $\alpha +1$ (the low energy gamma-ray
power law) and $1/3$, respectively, for the ETC. Although the ETC, GROCSE, and
GTOTE have varying integration times, only six of the 54 bursts in our sample
were shorter than five seconds (the exposure time of the ETC). These bursts are
slightly brighter in GROCSE and GTOTE. The difference is only a few tenths of a
magnitude though and the resulting distributions have the same maximum and
median brightness. The overall distributions are almost identical as well so
that Figure~3 applies to all three of these experiments. Figures 4a and~b show
similar distributions for {\it HETE}. 

\begin{figure}
\psfig{file=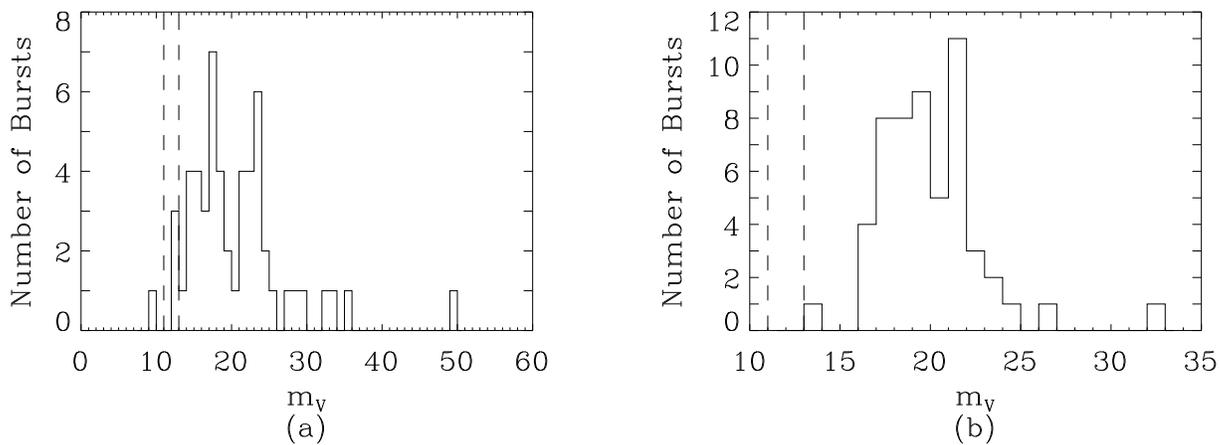}
\caption{The number of bursts at a given flux-based V magnitude for a 5~s
exposure assuming the flux extrapolates as a power law of index $\alpha+1$
(panel~a) or 1/3 (panel~b) from the gamma-rays. The dashed lines indicate the
sensitivities of the ETC ($m_V=11$) and GTOTE and GROCSE ($m_V=13$)
experiments.} 
\end{figure}
\begin{figure}
\psfig{file=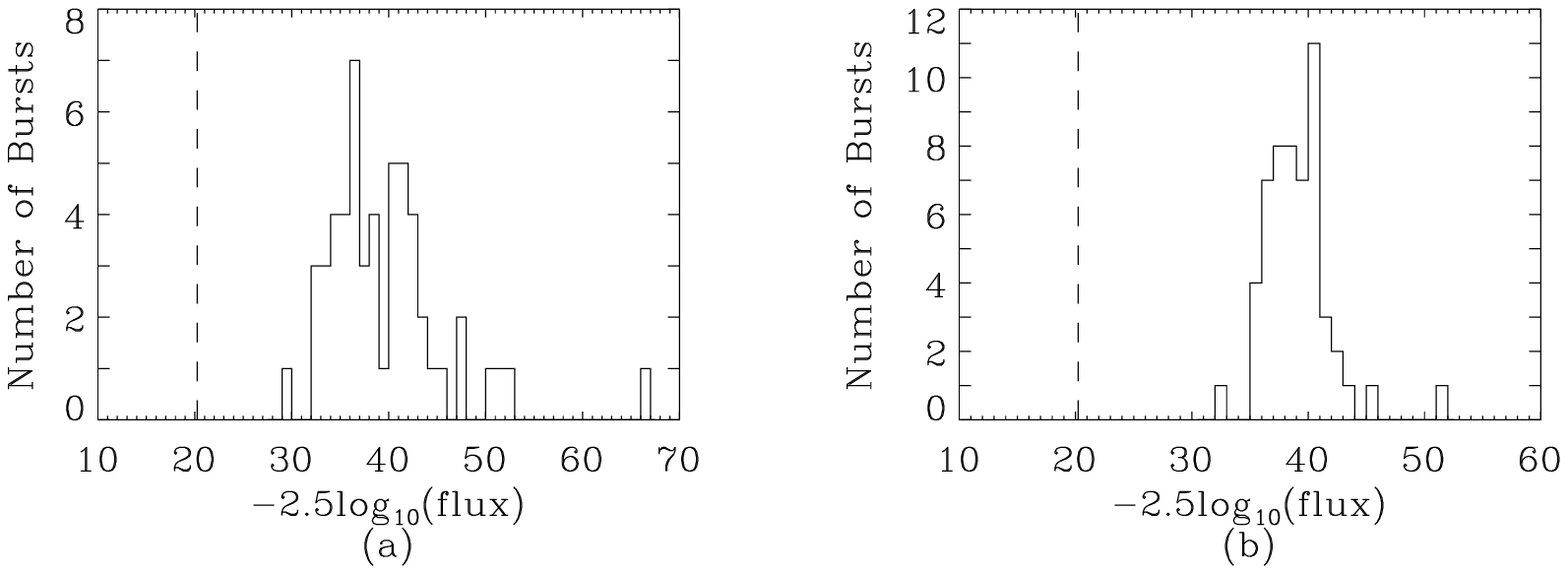}
\caption{The distributions of extrapolated fluxes for a 4s exposure in the
{\it HETE} band of 5-7~eV for energy power law index $\alpha+1$ (panel~a) or
1/3 (panel~b). The dashed line indicates {\it HETE}'s sensitivity.}
\end{figure}

An extrapolation with an energy power law index of~2 could correspond to the
Rayleigh-Jeans portion of a thermal spectrum.  For the Rayleigh-Jeans power law
to extend to 1~keV the radiating plasma should have a temperature above
$\sim1$~keV, which is possible in the extreme conditions expected in the burst
environment.  The thermonuclear model of Woosley \& Wallace (1982) predicted an
X-ray tail with a temperature greater than 2~keV; this model applied to local
bursts, and the predicted emission was not concurrent, but the model does show
that high temperatures are plausible. However, Table~3 shows that if this is
the only component of concurrent GRB emission at lower energies, GRBs will not
be observable at optical or UV wavelengths (assuming no afterglow). 

A power law index of 1/3 might result from low energy synchrotron emission
(Katz 1994; Tavani 1996a,b).  The low energy limit of the synchrotron spectrum
emitted by a single electron is a power law with an index of 1/3 below a
characteristic energy of $E_s = (3/4\pi)(e h B/m_e c) \gamma^2$, where $B$ is
the magnetic field, and $\gamma$ is the Lorentz factor corresponding to the
electron's energy.  The synchrotron power law indices of $\sim -1$ familiar
from radio astronomy result from a power law electron distribution.  If the
electron distribution has a low energy cutoff $\gamma_c$, then the low energy
single electron synchrotron power law of index 1/3 will be observed below
$E_s(\gamma_c)$.  Note that the energy spectral index (1+$\alpha$) of the low
energy component of the spectral function in eq.~(1) is typically of order 1/3.
In Figures 2, 3b, and~4b we present the optical-UV burst distributions for an
energy index of 1/3. For this case the brightest burst in our sample would have
appeared as a $m_V \sim 13$ transient, just at the sensitivity limit of GROCSE.
Therefore, we doubt current instruments can effectively test this model of
counterpart emission. 

The results of previous optical searches rule out some of the spectral indices
considered in this simple model. In particular, Greiner et~al.\ (1994) examined
sky patrol plates from several sites which were coincident with GRBs. Three
plates from the Ond\v rejov patrol were coincident with bursts in our sample.
These plates had four hour exposures with sensitivity to events brighter than
$m_V\sim$3. The three bursts which overlap with our sample (3B911127, 3B920525,
and 3B920530) would have produced optical images at least this bright if the
spectrum was a power law of index less than $-3$ below 1~keV.  (Published
before the 3B catalog was released, this result is still valid since the
difference in the positions of these bursts in the 2B and 3B catalogs is less
than the plate dimensions---J.~Greiner 1996, private communication) This rules
out more steeply falling power laws, which is not a severe constraint. 

Lacking compelling physical guidance for our expectations, the simplest
hypothesis is that spectra in the optical band have the same index as at higher
energies (the $\alpha +1$ case). If this is true, then the ETC, GROCSE, and
GTOTE have a chance of detecting the very brightest bursts which trigger BATSE
and are distributed by the BACODINE network. The ETC would have seen only the
brightest burst in the sample while GROCSE and GTOTE would have seen four. {\it
HETE} would not fare as well, being unable to detect any GRBs in the UV if the
$\alpha +1$ extrapolations hold. 

Being able to detect a GRB does not imply it will be observed. Assuming that
each site can image half the sky half the time (night only), then there is a
25\% chance a BATSE GRB can be slewed to.  For simplicity we neglect other
factors which reduce the duty cycle, such as inclement weather and moonlight.
Therefore, GROCSE and GTOTE might be
able to detect one burst per year for the $\alpha +1$ extrapolation and the ETC
might record a positive detection every four years. All of these detectors have
a monitor mode which could catch a burst BATSE does not see. BATSE has
$\sim$33\% sky coverage, so the number of potentially detectable GRBs is
tripled assuming a site can view all the sky all the time. However, 75\% of
these are lost to a given ground based site when earth blockage and daytime are
included. Most of the rest are inaccessible because the fields-of-view of these
instruments is small. The ETC has the largest field-of-view and it covers only
10\% of the sky accessible to it. Therefore, detection in the monitor mode of
any of these instruments is unlikely for this extrapolation. 

The instruments discussed here would have difficulty observing simultaneous
optical emission from a GRB if optical photons are generated as part of the
same spectrum as the gamma-rays. This implies that if concurrent emission is
observed more than about once a year, the energy flux must follow a power law
of index greater than $\alpha +1$, indicative of another spectral component
since GRB spectra are concave down above 10~keV. However, there are indications
of emission in excess of simple extrapolations below $\sim 10$~keV in 10\% of
GRBs (Preece et al.\ 1996). Therefore, aside from the discovery of a GRB
counterpart, the observation of simultaneous optical emission would indicate a
multifaceted emission process, providing new and important information about
the burst environment. 
\section{Summary}
Using spectra of 54 bright GRBs from BATSE, we have determined the
expected optical brightnesses for simultaneous optical emission for an
empirical extrapolation of the observed gamma-ray spectrum. We
find that in order for concurrent optical emission to be observable, the
spectral index of a power law extrapolation must be negative at long
wavelengths (more flux at lower energies). A simple extrapolation to optical
wavelengths is only marginally detectable, providing about one observable burst
per year for the GROCSE and GTOTE telescopes. This implies that if optical
transients concurrent with burst emission are observed more frequently, an
additional spectral component must exist. 

\acknowledgements

We are indebted to S. Barthelmy and R. Vanderspek for sharing details of their
instruments. We also thank R. Narayan for suggesting this project and
W.~Coburn, G.~Huszar, D.~Marsden, and L.~Peterson for their assistance. We
appreciate the insightful comments of the referee, J.~Greiner. This work was
supported by NASA contract NAS8-36081. 

\clearpage

% Table 1
%
\begin{deluxetable}{ccccccccc}
\scriptsize
\tablecolumns{9}
\tablewidth{0pc}
\tablecaption{Fluence Based Expected Brightness\tablenotemark{a}}
\tablehead{
\colhead{Index} & \colhead{min($m_V$)} & \colhead{median($m_V$)} &
\colhead{min($m_B$)} & \colhead{median($m_B$)} &
\colhead{min($m_U$)} & \colhead{median($m_U$)} &
\colhead{max($flux$)\tablenotemark{b}} & 
\colhead{median($flux$)\tablenotemark{b}} }
\startdata
$\alpha + 1$ & \phs\phn 7.8 & \phs 15.3 & \phs\phn 8.0 & \phs 15.4 &
\phs\phn 7.2 & 14.4 & \sci{1.03}{-11} & \sci{1.72}{-14} \nl
$-4$ & $-17.3$ & \phn $-12.3$ & $-16.1$ & $-11.1$ &
$-16.0$ & $-11.1$ & \sci{3.13}{-3\phn} & \sci{3.57}{-5\phn} \nl
$-3$ & $-10.7$ & \phn $-5.7$ & \phn $-9.7$ & \phn $-4.7$ &
$-10.0$ & $-5.0$ & \sci{1.81}{-5\phn} & \sci{2.06}{-7\phn} \nl
$-2$ & \phn $-4.1$ & \phs\phn 0.9 & \phn $-3.4$ & \phs\phn 1.6 &
\phn $-3.8$ & \phs\phn 1.2 & \sci{1.05}{-7\phn} & \sci{1.20}{-9\phn} \nl
$-1$ & \phs\phn 2.5 & \phs 7.5 & \phs\phn 3.0 & \phs\phn 8.0 &
\phs\phn 2.3 & \phs\phn 7.3 & \sci{6.22}{-10} & \sci{7.08}{-12} \nl
\phs 0 & \phs\phn 9.1 & \phs 14.1 & \phs\phn 9.3 & \phs 14.3 &
\phs\phn 8.4 & \phs 13.4 & \sci{3.70}{-12} & \sci{4.21}{-14} \nl
1/3 & \phs 11.3 & \phs 16.3 & \phs 11.4 & \phs 16.4 & \phs 10.4 &
\phs 15.4 & \sci{6.71}{-13} & \sci{7.64}{-15} \nl
\phs 1 & \phs 15.7 & \phs 20.7 & \phs 15.6 & \phs 20.6 &
\phs 14.5 & \phs 19.5 & \sci{2.22}{-14} & \sci{2.53}{-16} \nl
\phs 2 & \phs 22.3 & \phs 27.3 & \phs 22.0 & \phs 27.0 &
\phs 20.6 & \phs 25.6 & \sci{1.34}{-16} & \sci{1.52}{-18} \nl
\phs 3 & \phs 28.9 & \phs 33.9 & \phs 28.3 & \phs 33.3 &
\phs 26.7 & \phs 31.7 & \sci{8.21}{-19} & \sci{9.34}{-21} \nl
\phs 4 & \phs 35.5 & \phs 40.4 & \phs 34.6 & \phs 49.6 &
\phs 32.8 & \phs 37.8 & \sci{5.06}{-21} & \sci{5.76}{-23} \nl
\enddata
\tablenotetext{a}{Magnitudes are for bursts with all their fluence in 1s.}
\tablenotetext{b}{Values for {\it HETE} are ergs/cm$^2$-s in the $5-7$~eV band.}
\end{deluxetable}

%\clearpage

% Table 2
%
\begin{deluxetable}{cccc}
\footnotesize
\tablecaption{Instrument Summary}
\tablewidth{0pt}
\tablehead{
\colhead{Instrument} & \colhead{Field of} & \colhead{Exposure} &
\colhead{Sensitivity} \nl
\colhead{} & \colhead{View (sr)} & \colhead{Time} & \colhead{}
}
\startdata
Sonneberg\tablenotemark{c} & 0.15 & 40 m & 6 $(m_B)$\tablenotemark{a} \nl
Ond\v rejov\tablenotemark{c} & 4.36 & 4 hr & 3 $(m_V)$\tablenotemark{a} \nl
Odessa\tablenotemark{c} & 1.10 & 30 m & 6 $(m_B)$\tablenotemark{a} \nl
Dushanbe\tablenotemark{c} & 0.49 & 1 hr & 6 $(m_V)$\tablenotemark{a} \nl
ETC\tablenotemark{d} & 0.75 & 5 s & 11 $(m_V)$\tablenotemark{b} \nl
GROCSE\tablenotemark{e} & 0.621 & 0.4 s & 13 $(m_V)$\tablenotemark{b} \nl
GTOTE\tablenotemark{f} & $0.06-0.12$ & $3.3-4$ s &
$12-14$ ($m_V$)\tablenotemark{b} \nl
{\it HETE}\tablenotemark{g} & 1.7 & 4 s &
$8\times 10^{-9}$ ergs/cm$^2$-s ($5-7$~eV)\tablenotemark{b} \nl
\enddata
\tablenotetext{a}{Limiting magnitude for 1s flash.}
\tablenotetext{b}{Sensitivity to field stars.}
\tablenotetext{c}{Greiner et al.\ (1994)}
\tablenotetext{d}{Vanderspek et al.\ (1994)}
\tablenotetext{e}{Akerlof et al.\ (1994),
values for the second generation camera are used here}
\tablenotetext{f}{S. Barthelmy (1995), private communication (preliminary)}
\tablenotetext{g}{Ricker et al.\ (1992); R. Vanderspek (1995), private
communication}
\end{deluxetable}

%\clearpage

% Table 3
%
\begin{deluxetable}{ccccccccc}
\footnotesize
\tablecolumns{9}
\tablewidth{0pc}
\tablecaption{Expected Brightness of Bursts for Exposure Duration}
\tablehead{
\colhead{} & \multicolumn{2}{c}{ETC, GTOTE \& GROCSE} & \colhead{} &
\multicolumn{2}{c}{Ond\v rejov\tablenotemark{1}} & \colhead{} &
\multicolumn{2}{c}{{\it HETE}\tablenotemark{2}} \\
\cline{2-3} \cline{5-6} \cline{8-9} \\
\colhead{Index} & \colhead{min($m_V$)} & \colhead{median($m_V$)} & \colhead{} &
\colhead{min($m_V$)} & \colhead{median($m_V$)} & \colhead{} &
\colhead{max($flux$)} & \colhead{median($flux$)} }
\startdata
$\alpha + 1$ & \phs\phn 9.9 & \phs 19.2 & & \phs 18.2 & \phs 25.7 & &
\sci{1.57}{-12} & \sci{5.12}{-16} \nl
$-4$ & $-15.2$ & \phn $-8.8$ & & \phn $-6.9$ & \phn $-1.9$ & &
\sci{4.75}{-4\phn} & \sci{1.52}{-6\phn} \nl
$-3$ & \phn $-8.6$ & \phn $-2.2$ & & \phn $-0.3$ & \phs\phn 4.7 & &
\sci{2.75}{-6\phn} & \sci{8.82}{-9\phn} \nl
$-2$ & \phn $-2.0$ & \phs\phn 4.4 & & \phs\phn 6.3 & \phs 11.3 & &
\sci{1.60}{-8\phn} & \sci{5.14}{-11} \nl
$-1$ & \phs\phn 4.6 & \phs 11.0 & & \phs 12.9 & \phs 17.9 & &
\sci{9.44}{-11} & \sci{3.03}{-13} \nl
\phs 0 & \phs 11.2 & \phs 17.6 & & \phs 19.5 & \phs 24.5 & &
\sci{5.61}{-13} & \sci{1.80}{-15} \nl
\phs 1/3 & \phs 13.4 & \phs 19.8 & & \phs 18.7 & \phs 23.7 & &
\sci{1.02}{-13} & \sci{3.27}{-16} \nl
\phs 1 & \phs 17.8 & \phs 24.2 & & \phs 26.1 & \phs 31.1 & &
\sci{3.37}{-15} & \sci{1.08}{-17} \nl
\phs 2 & \phs 24.4 & \phs 30.8 & & \phs 32.7 & \phs 37.7 & &
\sci{2.04}{-17} & \sci{6.54}{-20} \nl
\phs 3 & \phs 30.9 & \phs 37.4 & & \phs 39.3 & \phs 44.3 & &
\sci{1.25}{-19} & \sci{4.00}{-22} \nl
\phs 4 & \phs 37.5 & \phs 44.0 & & \phs 45.9 & \phs 50.9 & &
\sci{7.68}{-22} & \sci{2.46}{-24} \nl
\enddata
\tablenotetext{1}{Values for Sonneberg, Odessa, and Dushanbe are typically
2 magnitudes brighter than for Ond\v rejov.}
\tablenotetext{2}{Values for {\it HETE} are in terms of ergs/cm$^2$-s in the
$5-7$~eV band.} 
\end{deluxetable}

\end{document}